# 基于 $\ell_p$ 范数漏溢式 LMS 的稀疏系统辨识

冯永[1], 曾瑞[2], 伍家松[2]

1. 东南大学生物科学与医学工程学院, 南京, 210096

2. 东南大学计算机科学与工程学院, 南京, 210096

**摘要**：漏溢式最小均方(leaky LMS, LLMS)算法用于解决标准 LMS 算法在处理系统输入信号高度相关时性能衰减的问题，但是它们均未考虑利用系统冲击响应的稀疏性来进一步提高算法性能。因此本文提出一种基于 $\ell_p$ 范数约束的 LLMS 的新算法($\ell_p$-LLMS)，作为标准 LLMS 算法的延伸，把 $\ell_p$ 范数约束引入到 LLMS 算法的代价函数中，从而加强稀疏系统的辨识效果。计算机仿真实验表明，对于稀疏系统辨识，在不同稀疏度条件下，该算法在稳态误差方面均优于标准 LMS, LLMS 和 $\ell_p$-LMS 算法.

**关键词**：自适应滤波；$\ell_p$ 范数约束；稀疏系统辨识；漏溢式最小均方；最小均方算法

# *p* Norm Constraint Leaky LMS Algorithm for Sparse System Identification

*FENG Yong[1], ZENG Rui[2], WU Jiasong[2]*

1. School of Biological Science and Medical Engineering, Southeast University, Nanjing 210096, China

2. School of Computer Science and Engineering, Southeast University, Nanjing 210096, China

**Abstract:** This paper proposes a new leaky least mean square (leaky LMS, LLMS) algorithm in which a norm penalty is introduced to force the solution to be sparse in the application of system identification. The leaky LMS algorithm is derived because the performance of the standard LMS algorithm deteriorates when the input is highly correlated. However, both of them do not take the sparsity information into account to yield better behaviors. As a modification of the LLMS algorithm, the proposed algorithm, named $\ell_p$-LLMS, incorporates a *p* norm penalty into the cost function of the LLMS to obtain a shrinkage in the weight update equation, which then enhances the performance of the filter in system identification settings, especially when the impulse response is sparse. The simulation results verify that the proposed algorithm improves the performance of the filter in sparse system settings in the presence of noisy input signals.

## Introduction

In practice of communication, it is very common to encounter a sparse system whose impulse response contains very few nonzero coefficients, while other taps are zeros or nearly zeros. For example, sparse sparse acoustic echo path, digital TV transmission channels, sparse wireless multi-path channels, etc.. The least mean square (LMS) algorithm is the most widely-used algorithm among the traditional adaptive filtering algorithms for system identification, due to its computational simplicity, efficiency and robustness [1]. However, that it does not assume any structural information about the system to be identified, in turn, results a poor performance in terms of the steady-state excess mean square error (EMSE) as well as the convergence speed [2]. Moreover, the performance of the standard LMS deteriorates if the input of the system is highly correlated. Thus, some modifications of the LMS have been proposed to overcome this problem, and the leaky LMS (LLMS) algorithm [3] is among the famous LMS variants.

Recently there emerge some sparse LMS algorithms with different norm constraints, for example, the $\ell_1$-norm penalty LMS ($\ell_1$-LMS) [2, 4], $\ell_0$-norm penalty LMS ($\ell_0$-LMS) [5, 6] and $\ell_p$-norm penalty LMS ($\ell_p$-LMS) [7, 8], in which the corresponding $\ell_1$, $\ell_0$ and $\ell_p$ norms are incorporated into the cost function of the standard LMS algorithm, respectively, to increase the convergence speed and decrease the mean square error (MSE) simultaneously.

To combine the advantages of both the sparse LMS and LLMS algorithm, we propose a $p$ norm constraint LLMS algorithm, in which a $p$-norm penalty is introduced into the cost function of the LLMS to force the solution to be sparse in the application of system identification.

This paper is organized as follows. In Section 2, the standard LMS and leaky LMS algorithm are briefly reviewed and then the proposed algorithm is derived, following by the simulation results that compare the performance of the proposed algorithm with those of the standard leaky-LMS, standard LMS and $\ell_p$-LMS algorithm in sparse system identification settings in Section 3. Finally, in section 5, the conclusion are presented.

## Algorithms

Let $y_k$ be the output of an unknown system with an additional noise $n_k$ at time $k$, which can be written as

$$y_k = \mathbf{w}^T \mathbf{x}_k + n_k, \tag{1}$$

where the weight $\mathbf{w}$ of length $N$ denotes the impulse response of the unknown system, $\mathbf{x}_k = [x_k, x_{k-1}, \cdots, x_{k-N+1}]^T$ represents the input vector with covariance $\mathbf{R}$, and $n_k$ is a stationary noise with zero mean and variance $\sigma_k^2$. Given the input $\mathbf{x}_k$ and output $y_k$ following the aforementioned settings, the LMS algorithm has been proposed to estimate the weight vector $\mathbf{w}$. The cost function $J_k$ of the standard LMS algorithm is defined as

$$J_k = e_k^2 / 2, \tag{2}$$

where $e_k = y_k - \mathbf{w}_k^T \mathbf{x}_k$ denotes the instantaneous error and $\mathbf{w}_k = [[\mathbf{w}_k]_1, [\mathbf{w}_k]_2, ..., [\mathbf{w}_k]_N]^T$ represents the estimated weight of the system at time $k$, while "1/2" is applied for the convenience of computation. Thus, the update equation is then written as

$$\mathbf{w}_{k+1} = \mathbf{w}_k - \mu \frac{\partial J_k}{\partial \mathbf{w}_k} = \mathbf{w}_k + \mu e_k \mathbf{x}_k, \tag{3}$$

where $\mu$ is the step size which satisfies $0 < \mu < \lambda_{max}^{-1}$ with $\lambda_{max}$ being the maximum eigenvalue of $\mathbf{R}$.

In order to mitigate the weight drift problem in the LMS algorithm under the condition that the input signal is highly correlated in the application of system identification, the LLMS algorithm is introduced, which improves the convergence and stability. The LLMS uses a leakage factor $\gamma (0 < \gamma < 1)$ to control the weight update of the LMS algorithm, and its cost function is defined as

$$J_k = e_k^2 / 2 + \gamma \|\mathbf{w}_k\|_2^2 \qquad (4)$$

And the weight vector is then updated by

$$\mathbf{w}_{k+1} = (1 - \mu\gamma)\mathbf{w}_k + \mu e_k \mathbf{x}_k \qquad (5)$$

For a sparse system in which most of the taps in the weight are exactly or nearly zeros, the $\ell_p$-LMS algorithm has been proposed with the new cost function $J_{k,p}$ represented by

$$J_{k,p} = e_k^2 / 2 + \gamma_p \|\mathbf{w}_k\|_p, \qquad (6)$$

where the $p$ norm is defined as $\|\mathbf{w}_k\|_p \triangleq \left( \sum_{i=1}^{N} |[\mathbf{w}_k]_i|^p \right)^{1/p}$ with $0 < p < 1$, and $\gamma_p$ is a constant controlling the trade-off between the convergence rate and estimation error. Thus, the update of the $\ell_p$-LMS is derived as

$$\mathbf{w}_{k+1} = \mathbf{w}_k + \mu e_k \mathbf{x}_k - \rho_p \frac{\|\mathbf{w}_k\|_p^{1-p} \operatorname{sgn}(\mathbf{w}_k)}{\varepsilon_p + |\mathbf{w}_k|^{1-p}}, \qquad (7)$$

where $\rho_p = \mu\gamma_p$ weighting the $p$-norm constraint, $\varepsilon_p$ is a constant bounding the term and sgn($x$) is the sign function, which is 0 for $x = 0$, 1 for $x > 0$ and -1 for $x < 0$.

As an exploration, we take the idea of $p$ norm constraint in the LLMS, to obtain a new algorithm named the $\ell_p$-LLMS, whose cost function is defined as

$$J_k = e_k^2 / 2 + \gamma \|\mathbf{w}_k\|_2^2 + \gamma_p \|\mathbf{w}_k\|_p. \qquad (8)$$

Using gradient descent, the update equation for the $\ell_p$-LLMS is given by

$$\mathbf{w}_{k+1} = (1 + \mu\gamma)\mathbf{w}_k + \mu e_k \mathbf{x}_k - \rho_p \frac{\|\mathbf{w}_k\|_p^{1-p} \operatorname{sgn}(\mathbf{w}_k)}{\varepsilon_p + |\mathbf{w}_k|^{1-p}}. \qquad (9)$$

Note that we employ (1 + $\mu\gamma$) instead of (1 - $\mu\gamma$) here to obtain a lower stable error. Thus, it is not directly extended from the LLMS. It should also be noted that, similar to the cost function of the $\ell_p$-LMS, the cost function of the $\ell_p$-LLMS is not convex and therefore, the analysis of the global convergence and consistency of the corresponding algorithm is problematic. However, as it will be shown in the next section that the proposed $\ell_p$-LLMS algorithm outperforms the standard LMS, LLMS and the $\ell_p$-LMS in condition that the input signal is highly correlated in the sparse system identification settings.

## Simulations

In this section, the performance of the $\ell_p$-LLMS, measured in terms of the mean square deviation (MSD, defined as $\mathrm{MSD}_k = \mathrm{E}[\|\mathbf{w} - \mathbf{w}_k\|_2^2]$ ), is compared with those of the standard LMS, LLMS and $\ell_p$-LMS algorithm in the numerical simulations, which are aimed at showing their tracking and steady-state performances in the sparse system identification settings.

We estimate a sparse unknown system of 16 taps with 1, 4 or 8 taps that are assumed to be nonzeros, which makes the sparsity ratio (SR) be 1/16, 4/16 or 8/16, respectively. The positions of nonzero taps are chosen randomly and the values are 1's or -1's randomly. Initially, we set the SR = 1/16 before the 8000$th$ iteration, and after that, we have SR = 4/16 for the next 8000 iterations, and then SR = 8/16 for the last 8000 iterations, leaving a semi-sparse system. The input is a correlated signal which is generated by $x_{k+1} = 0.8 x_k + u_k$ and then normalized to variance 1, where $u_k$ is a white Gaussian noise with variance $10^{-3}$. The observed noise is assumed to be white Gaussian processes of length 8016 with zero mean and variance $10^{-2}$, i.e., the signal noise ratio (SNR) is set to be 20 dB. Other parameters are carefully selected as listed in Table 1. All the simulations are averaged over 200 independent runs to smooth out their MSD curves.

Table 1. Parameters of the algorithms in the experiment.

| Algorithms | $\mu$ | $\rho$ | $\varepsilon$ | $p$ | $\gamma$ |
|---|---|---|---|---|---|
| LMS | | / | / | / | / |
| $\ell_p$-LMS | | 0.0005* | | | / |
| leaky LMS | 0.015 | 0.0002** | 10 | 0.5 | 0.005 |
| $\ell_p$-LLMS | | 0.0001*** | | | |

*,**,*** are for SR=1/16, 4/16 and 8/16, respectively.

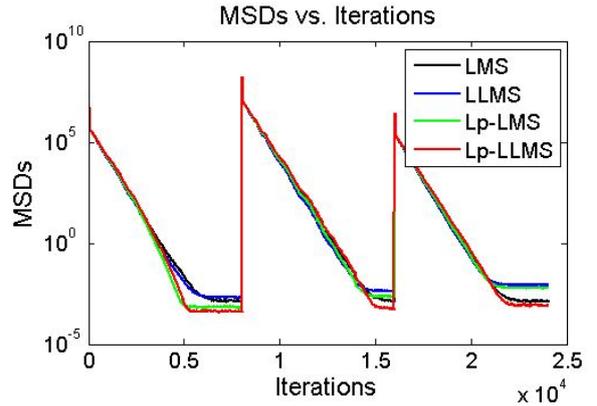

Fig. 1. MSD curves of different algorithms with different SRs

Fig. 1 shows the MSD curves of the algorithms tested for different sparsity levels, that is, SR = 1/6 for the first 8000 iterations, SR = 4/16 for the 8001$th$ to 16000$th$ iteration and SR=8/16 for the next 8000 iterations left. From Fig. 1, it can be seen that, for all different sparsity levels, i.e., SR = 1/16, 4/16 and 8/16, the proposed $\ell_p$-LLMS always achieves the best performance in term of steady-state MSD, followed by the $\ell_p$-LMS which convergences a little faster in the very sparse case. Moreover, when a system is very sparse, i.e., SR = 1/16, both the $\ell_p$-LMS and $\ell_p$-LLMS yield lower stable MSD than those of the LMS and LLMS, due to the function of the $p$-norm sparse constraint which attracts the taps of the impulse response to zeros. However, as the sparsity ratio increases, the $\ell_p$-LMS performs worse than the standard LMS algorithm, which is expected, since the $\ell_p$-LMS is developed to deal with sparse cases. But our $\ell_p$-LLMS still performs the best, owning to its feature which combines the advantages of both the standard LLMS and the $\ell_p$-LMS. Overall, our proposed $\ell_p$-LLMS outperforms the standard LMS, LMS and $\ell_p$-LMS for different sparsity levels of the system in sparse system identification settings.

# Conclusion

We propose the $\ell_p$-LMS algorithm which incorporates a $p$ norm constraint into the cost function of the leaky LMS algorithm to force the solution to be sparse in the application of system identification, thus it combines the advantages of both the leaky LMS and the $\ell_p$-LMS to achieve better performance while the input of the system is highly correlated. And the simulations show that our proposed $\ell_p$-LLMS outperforms the standard LMS, LLMS and $\ell_p$-LMS for different sparsity levels of the system in sparse system identification settings in the presence of noisy input signals.

Our future work will focus on the choices of the parameters of the $\ell_p$-LLMS and $\ell_p$-LMS algorithm for sparse system identification, for we found that the parameters leaky factor $\gamma$, $p$-norm trade-off $\rho$ and SNR as well as $p$ of these two $p$ norm constraint (L)LMS have significant effects on the performance of convergence rate as well as stable error, thus we will explore their relationships while also considering the sparsity level and step size.